\documentclass[preprint]{vgtc}
\ifpdf%                                % if we use pdflatex
  \pdfoutput=1\relax                   % create PDFs from pdfLaTeX
  \usepackage{graphicx}                % allow us to embed graphics files
  \DeclareGraphicsExtensions{.pdf,.png,.jpg,.jpeg} % for pdflatex we expect .pdf, .png, or .jpg files
\else%                                 % else we use pure latex
  \usepackage{graphicx}                % allow us to embed graphics files
  \DeclareGraphicsExtensions{.eps}     % for pure latex we expect eps files
\fi%

\graphicspath{{figures/}{pictures/}{images/}{./}} % where to search for the images

\usepackage{microtype}                 % use micro-typography (slightly more compact, better to read)
\PassOptionsToPackage{warn}{textcomp}  % to address font issues with \textrightarrow
\usepackage{textcomp}                  % use better special symbols
\usepackage{mathptmx}                  % use matching math font
\usepackage{times}                     % we use Times as the main font
\usepackage{cite}                      % needed to automatically sort the references
\usepackage{tabu}                      % only used for the table example
\usepackage{booktabs}                  % only used for the table example
\usepackage{wrapfig}
\usepackage{endnotes}

\newcommand{\etal}{et al.\@ }
\newcommand{\etals}{et al.\@'s }

\newcommand{\termcoin}[1]{\emph{#1}}

\newcommand{\parahead}[1]{\altparahead{#1.}}

\newcommand{\altparahead}[1]
{%
    \vspace{0.07in}%
    \noindent%
    \textbf{\textit{#1}}%
}
\newcommand{\secref}[1]{\hyperref[#1]{Sec.~\ref*{#1}}}
\newcommand{\noref}[1]{\hyperref[#1]{~\ref*{#1}}}
\newcommand{\appendixref}[1]{\hyperref[#1]{Appendix.~\ref*{#1}}}
\newcommand{\figref}[1]{\hyperref[#1]{Fig.~\ref*{#1}}}
\newcommand{\eqnref}[1]{\hyperref[#1]{Eqn.~\ref*{#1}}}
\newcommand{\tabref}[1]{\hyperref[#1]{Table ~\ref*{#1}}}

\onlineid{0}

\vgtccategory{Research}

\vgtcinsertpkg

\preprinttext{This manuscript was  presented at alt.VIS, a workshop co-located with IEEE VIS 2021 (held virtually)}
\title{Visualization for Villainy}

\author{Andrew M. McNutt\thanks{e-mail: mcnutt@uchicago.edu}\\ %
        \scriptsize University of Chicago %
\and Lilian Huang\thanks{e-mail: lilianhj@uchicago.edu}\\ %
     \scriptsize NORC at the University of Chicago %
\and Kathryn Koenig\thanks{e-mail: koenig1@uchicago.edu}\\ %
     \scriptsize Chicago Transit Authority}

\abstract{
Visualization has long been seen as a dependable and trustworthy tool for carrying out analysis and communication tasks---a view reinforced by the growing interest in applying it to socially positive ends. However, despite the benign light in which visualization is usually perceived, it carries the potential to do harm to people, places, concepts, and things. In this paper, we capitalize on this negative potential to serve an underrepresented (but technologically engaged) group: villains. To achieve these ends, we introduce a design space for this type of graphical violence, which allows us to unify prior work on deceptive visualization with novel data-driven dastardly deeds, such as emotional spear phishing and unsafe data physicalization. By charting this vile charting landscape, we open new doors to collaboration with terrifying domain experts, and hopefully, make the world just a bit worse.
} % end of abstract

\CCScatlist{
  Deceptive Visualization; Data Physicalization; Block-chain; Supervillainy---Traditional and modern; Evil; Harm Optimization;
}

\begin{document}


\firstsection{Introduction}

\maketitle

Most evil done with visualization today, just like with many other domains\cite{arendt1964eichmann}, is banal. The dashboards, spreadsheets, and reports that people make as part of their jobs are simply tools to design and carry out tasks. While the charts, graphs, and maps comprising these entities may serve a greater sinister purpose, their primary goal is seldom to do evil.
Therefore, this paper will not mention admirable endeavors such as Palantir's application of tools like machine learning and visual analytics\cite{wright2009palantir} to drive America towards becoming a surveillance police state\cite{winston2018palantir}. Neither will it mention Tableau contracting with ICE\cite{Tableau19Drawing}, an organization that has actively made life worse for countless vulnerable people.
We omit these because, while they have effectively supported the natural and reasonable goals of making the world worse, the visualizations themselves are mundane; the medium of visualization itself has not been honed and exploited to unleash its maximum potential for malice. Such pedestrian acts of villainy are beneath us as scholars of evil.

We instead explore the ways in which intentional harm can be brought upon viewers of visualization---that is, we seek to understand how we might better use visualization for overt villainy.
In doing so, we aim to open a new visualization frontier, in which evil is not incidental, but is foregrounded throughout every step of visualization practice---allowing cruelty to be, in fact, the point\cite{serwer2021cruelty}.

We believe this is a critical juncture to carry out this sinister work.
Recent efforts to use visualization for social good\cite{vis4good, vis4socialgood, data4change}
and the emerging thread of research focused on the ethical issues facing data visualization practitioners\cite{ doNoHarm, correll2019ethical, ehmeltopography} suggest a growing interest in using visualization for the benefit of marginalized people\cite{peck2019data, d2020data}, disabled people\cite{marriott2021inclusive, lundgard2019sociotechnical, wu2021understanding, Kim21Accessible}, and people in general.

We therefore hold that there is an urgent need to intervene now and provide countermeasures to these do-gooder efforts.

We must not be content to rest on our laurels and merely reflect on how visualization was a favored propaganda tool of the Nazis\cite{correll2019ethical}, whose terrifying assertions were swaddled in the seemingly objective rhetorical mode that visualizations carry (in effect deifying their worldview). We instead seek to forge ahead and provide cutting-edge, practical tools for sowing damage, despair, and distrust in the contemporary era.
We work towards these ends by developing a design space for villainous visualization within which we situate a number of established evil tactics and identify several new ones.
We build upon the works of scholars of applied graphical evil, such as Snider\cite{Snider11Evil} (who defined a number of sinister graphical forms), Correll \cite{correll2017black} (who described a family of black hat visualization attacks), and Tomlinson \cite{tomlinson2020suffering} (who described a family of design principles to center---but unfortunately not increase---suffering as part of the design process).
This work focuses the malicious intentions of the venerable CHI4Evil workshop\cite{soden2019chi4evil} upon the visualization sphere.
It is our hope that by carrying out this work, we will both open the door to collaboration with the sorts of villains who are typically excluded from visualization research, and enrich our partnerships with those normally included.

\begin{figure}
  \vspace{-0.4in}
  \centering
  \includegraphics[width=\linewidth]{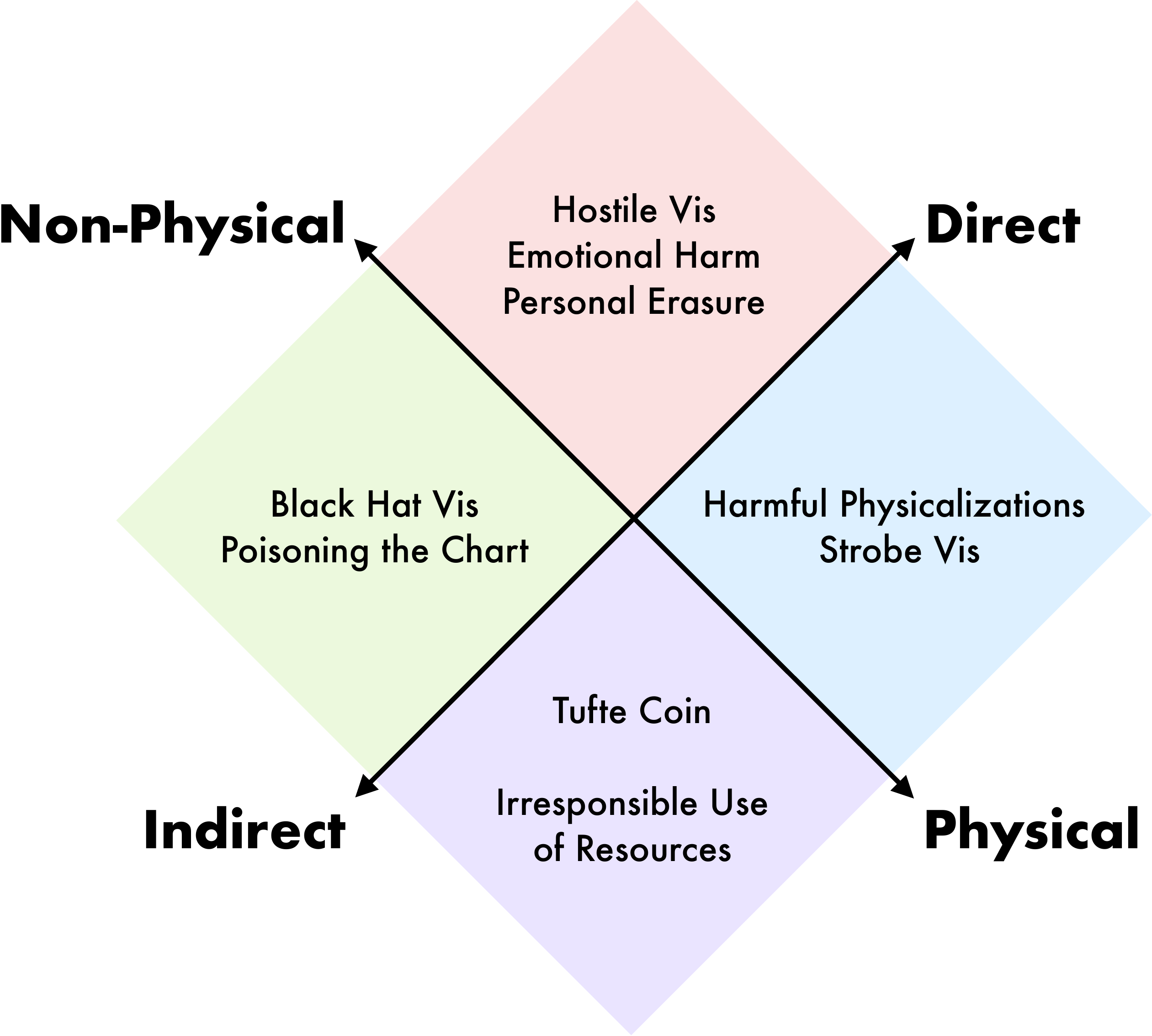}
  \caption{The design space of harm that visualizations can do.}
  \label{fig:alignment}
  %   \vspace{-0.2in}
\end{figure}

\begin{figure*}[t]
  \centering
  \includegraphics[width=0.9\linewidth]{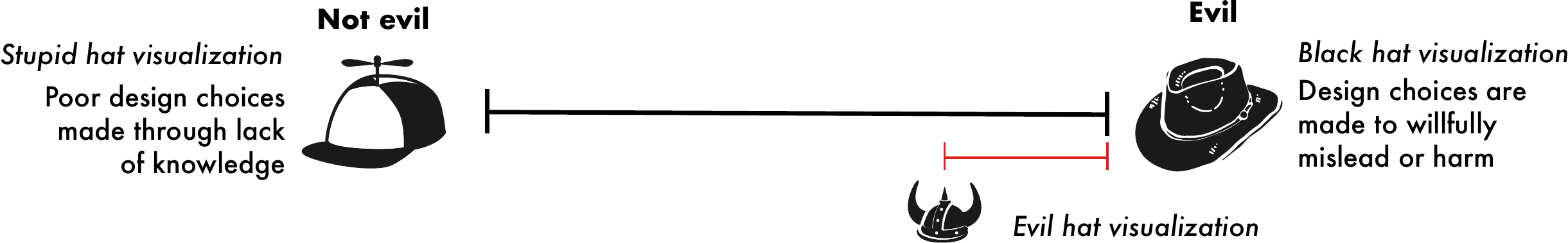}
  \caption{Bad visualizations come in many forms.
    Some may mislead because of unfortunate design choices made from a place of sincerity---such as the infamous ``Gun deaths in Florida'' chart\cite{Lallanilla14Misleading}---while others are made with a more direct intention to harm.
  }
  \label{fig:teaser}
  \vspace{-0.2in}
\end{figure*}

\section{Design space}

There are many ways in which one might achieve sinister ends using visualizations. For instance, one may create evil by handling sensitive data in a brash or offensive manner \cite{ehmeltopography}, or by using charts to create false impressions about government programs through intentionally confusing design choices\cite{NareaWarOnCharts}.
Given this variation, and in the interest of bringing a greater host of evils into the world, we create a design space that might unify these various evil possibilities.
We show our design space in \figref{fig:alignment}, categorize several past mechanisms of malice within it, and introduce a number of novel tactics.
This design space is formed by taking the non-malfeasance expressed in the oft-misquoted Hippocratic oath (``first do no harm'')\footnote{Coincidentally, the Urban Institute recently released a report recommending that practitioners strive to ``Do No Harm''\cite{doNoHarm} to vulnerable communities through the design of their visualizations.} as our ethical departure point.
We define our villainy in opposition to this injunction: \termcoin{First, Let Us Do Harm}.

Following this axiom, we partition the space of possible harms in twain, twice.
First, we note that harm can be either physical or non-physical; second, we partition by impact: are the wrongs wrought directly (aimed at the viewer themselves) or indirectly (aimed at the environment within which the viewer exists, in such a way that harm trickles down to them)?
This model takes inspiration from the tabular form of the matrix of domination\cite{collins2002black}, but rather than identifying general venues in which to carry out structural oppression (which is itself a worthy goal), we instead seek specific ways that visualizations can be operationalized to do harm.
We select these dimensions from the infinite space of possible insidious ingredients, not because they perfectly capture the entire evil experience---but because they  allow us a useful vantage point from which to consider visualization villainy.
Box famously noted that ``all models are wrong''\cite{box1979robustness}, but had he seen our model---which makes even malicious models, such as those enacting deep learning for phrenology\cite{Ongweso20Phrenology}, seem sanctimonious in comparison---he would have likely reappraised that some models are alright.
The remainder of the paper will be a tour of these terrors, describing how various evil ends might be enacted in each of these categories.

\subsection{Image Control (Non-Physical Indirect)} Data visualizations are principally focused on communication, and thus the most commonly practiced strains of evil involve manipulating the understanding that the reader gains from viewing a visualization. There are countless visualizations that communicate their message poorly\cite{WTFVis}, or unintentionally misinform the reader (what might be called ``stupid hat'' visualizations)---as noted in \figref{fig:teaser}.
In contrast, here we focus on charts whose form is intentionally used to create harm through miscommunication and misinformation.

\begin{figure}
  \centering
  \includegraphics[width=\linewidth]{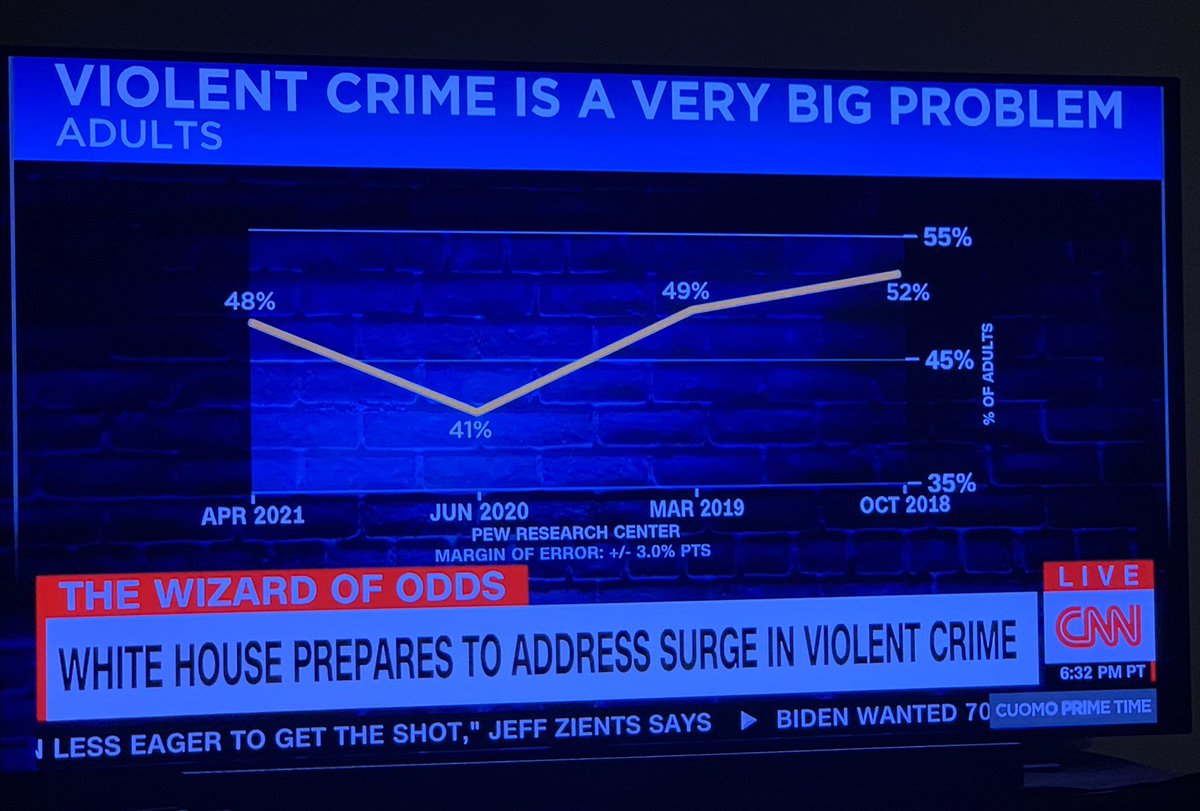}
  \caption{This graphic skilfully deceives by (among other tactics) reversing the x-axis direction to falsely imply a larger effect.\cite{badCNNGraph}
  }
  \label{fig:badaxes}
  \vspace{-0.2in}
\end{figure}

\parahead{Black Hat Visualization}
We begin with the most commonplace of our assaults, which intentionally misuses the form of a visualization to give a false impression.
Correll and Heer \cite{correll2017black} usefully describe a family of black hat visualizations, which are typically ``man in the middle'' attacks. In these attacks, a malicious designer manipulates a chart in such a way as to obscure or obfuscate the data, in order to present their own preferred message. Tactics include breaking conventions, nudging, and the use of non-sequitur visualizations (which appear to encode data as charts, but in fact merely use them as decoration).
Pandey \etal \cite{pandey2015deceptive} describe a series of attacks related to truncated and inverted axes (as in \figref{fig:badaxes}), aspect ratios, and area encodings.
Woodin \etal \cite{woodin2021conceptual} explore the deceptive potential of inverted axes in the context of metaphor.
% https://imgflip.com/memegenerator/239304046/Joey-seeing-himself-on-TV
McNutt \etal \cite{mcnutt2020surfacing} describe a wide family of errors that can be forced upon users from across the visualization pipeline, to cause what they term \emph{visualization mirages}.
Lauer and O'Brien \cite{lauer2020people} describe and demonstrate the deceptive power of a variety of misleading tactics.

\begin{figure}
  \centering
  \includegraphics[width=0.9\linewidth]{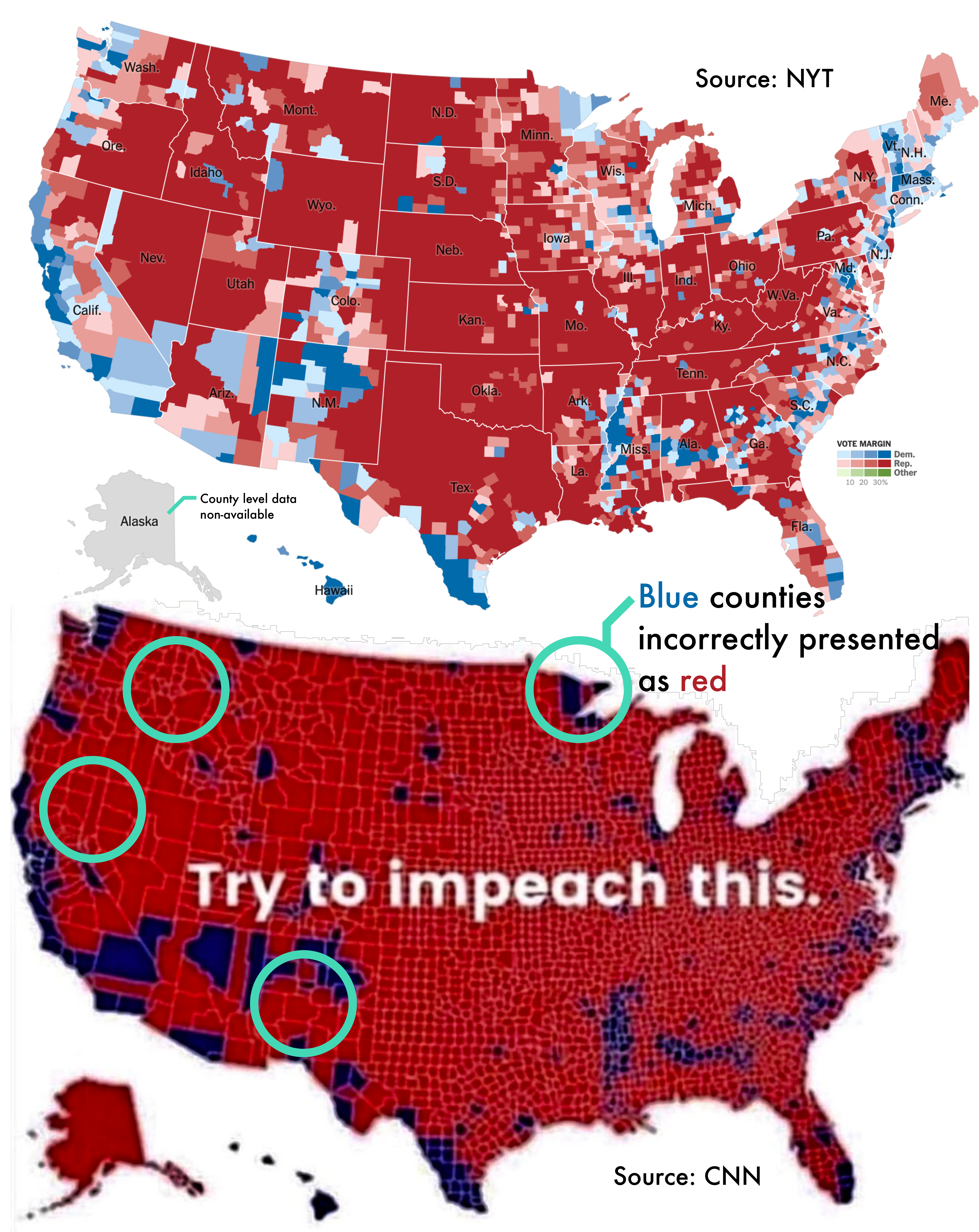}
  \caption{The viral ``Impeach This'' chart cunningly stacks common deceptions (conflating ranges as binaries, ``land doesn't vote''), masking more devious data manipulations. Sources \cite{Lybrand19Trump, NYT2016Map}.}
  \label{fig:impeachthis}
  \vspace{-0.2in}
\end{figure}

When readers assume that the information they are given is correct, there is ample room to distort, cherry-pick, or simply change the data. Robinson explores the space of viral visualizations and maps \cite{robinson2019elements}, and the way that they can propagate and disseminate false information, which offers an intriguing and high-impact way to sow chaos. The widely-circulated ``Impeach This'' map exemplifies this strategy. This viral visualization ostensibly shows a county-level choropleth of the 2016 United States presidential election, colored red or blue based on the winner of the county. In addition to exemplifying the \emph{land doesn't vote} mirage\cite{mcnutt2020surfacing}, the version of this graphic most prominently available features several data corruptions, rendering ``multiple blue counties won by Hillary Clinton as red counties won by Trump.''\cite{Lybrand19Trump}---as seen in \figref{fig:impeachthis}.
In this vein, Pavliuc and Dykes \cite{pavliuc2020designing} use network visualizations to celebrate several state-based disinformation campaigns.

This tactic derives its power from the fact that visualizations are often understood as being objective depictions of the data, and are not recognized as the rhetorical communications\cite{hullman2011visualization} that they truly are.
The moralist La Rochefoucauld notes ``truth does not do as much good in the world as the appearance of it does evil''\cite{de2007collected}; by swaddling ourselves in the gauze of faux-objectivity carried by charts, we have ample room to deliberately mislead and misinform.

% \am{``We cannot let maps, as a medium for communicating information, be co-opted by people with nefarious intentions. I pledge to do my part by clearly noting if a map is a parody in the future.'' \url{https://www.forbes.com/sites/brianbrettschneider/2018/11/23/lessons-from-posting-a-fake-map/?sh=3a6d176259ec}}

\parahead{Poisoning the Chart} The assumption of the unassailable objectivity of visualization has great utility; however, confusion and dissent can also be invoked by piercing this veil. ``Poisoning the Well'' is a well known argumentative fallacy\cite{walton2006poisoning} in which doubt is sown against a speaker by undermining their credibility, often by presenting information casting them in a negative light, regardless of whether or not said information is true (i.e. accusing them of some bullshit\cite{frankfurt2009bullshit}). For instance, consider a situation in which Bob tells you the water in a well is not poisonous. Alice comes along and tells you that Bob is a liar, has recently poisoned several puppies, and is guilty of tax fraud. Even if you do not believe Alice, you might find yourself disinclined to sample the water.
This line of attack can be usefully applied to visualization by planting a seed of doubt in the medium itself, the chart makers, or even the data, thereby \termcoin{Poisoning the Chart}.
% my pie saying "does not contain spiders" is raising a lot of questions already answered by the words on my pie

Once a viewer is made aware that a single deception has taken place (even if it is brief and for a purpose), they are less likely to trust any other information held by that visualization \cite{ritchie2019lie}.
There are many ways this might be achieved, such as annotations to careful misuse of the anchoring effect\cite{mcnutt2020surfacing}.
Yet such elaborate strategies may not even be necessary, as a well-placed strong-man can simply sharpie over a perfectly normal visualization and assert that their chosen conclusions are true, thus capitalizing on political polarization to create an air of uncertainty and confusion.
Lee \etal \cite{lee2021viral} document the development of a culture of visual analytics among a particularly doubt-ridden group (anti-maskers), and highlight how mistrust of the establishment can generate public fervor---and, we note, even death under some fortuitous circumstances.

Beyond sowing doubt, one might poison a chart by causing interaction with it to be perceived as undesirable. For instance, this might be carried out by engaging in aggressive patenting, such that public perception of a chart form is tainted by the turbid machinery of the legal system. One might simply patent several dozen commonly understood ideas, visualization techniques, or chart forms, and then publicly bring suit against prominent practitioners. This is likely to decrease any interest in using that chart, and may even foment distrust in prior usages of it.

\begin{figure}[t]
  \centering
  \includegraphics[width=\linewidth]{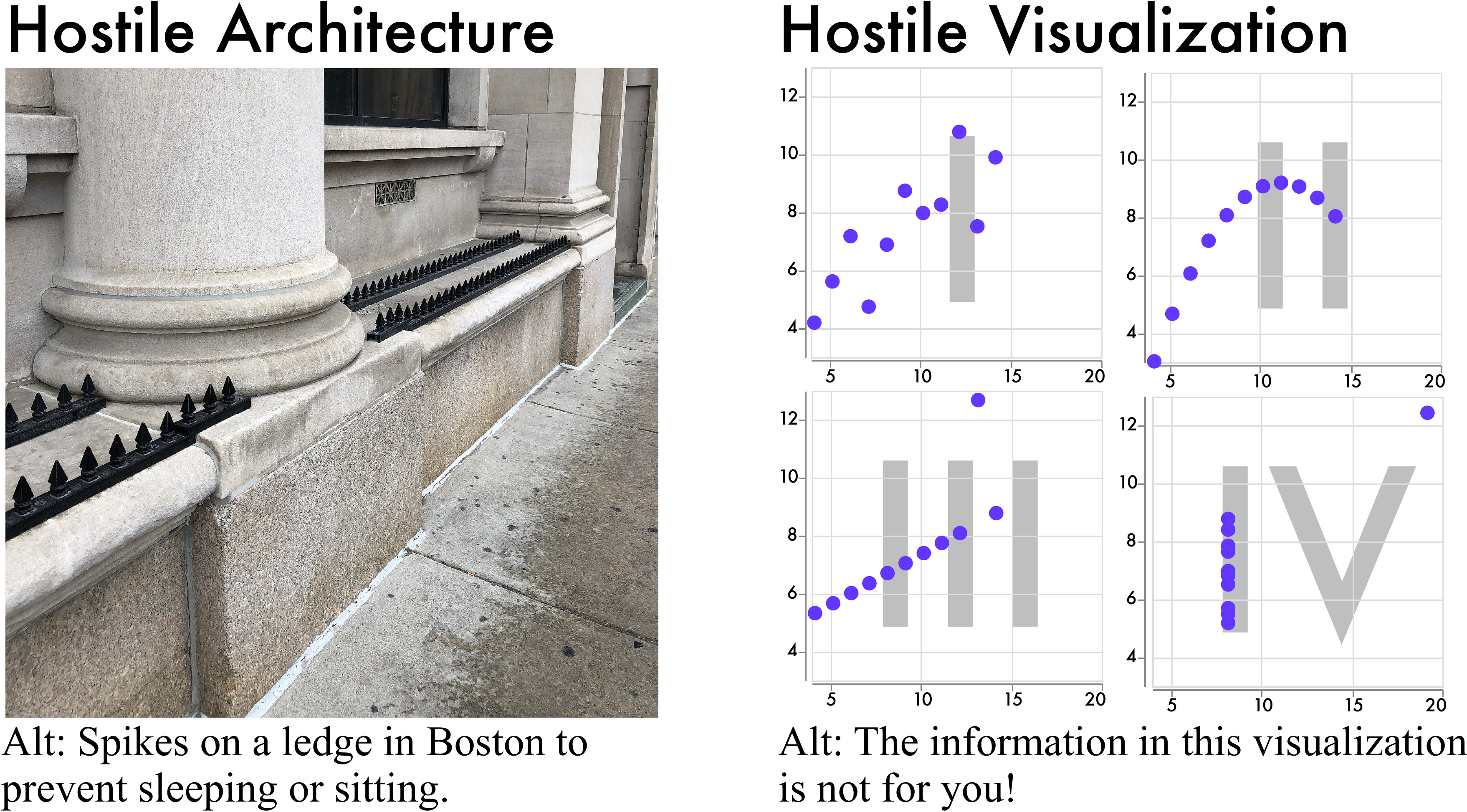}
  \vspace{-0.3in}
  \caption{A common hostile architecture technique (placing spikes where undesirables might rest) and a proposed hostile visualization technique (replacing descriptive alt text with antagonistic messages).}
  \label{fig:hostile}
  \vspace{-0.2in}
\end{figure}

\subsection{Feeling Personally Attacked (Non-Physical Direct)}

While it is reasonable to characterize all viewers as white-cis-able-bodied-young-wealthy-urban-educated-Americans (as many visualizations do), sometimes individuals will audaciously exhibit identities departing from this natural norm. Here, we consider the ways in which these deviant characteristics might be hijacked for harm.

\parahead{Hostile Visualization}
Many visualizations are inaccessible not on purpose, but by accident:
the designer, ignorant of accessibility guidelines, makes decisions that render their visualizations difficult or impossible to parse by viewers with visual impairments.
While such unintentional hostilities are appreciated, we propose taking this further and making these values explicit and deliberate.
To accommodate this intention, we yoke together two unrelated fields of design.
Chivukula \etal describe artifacts of asshole design as having ``clear malicious or deceptive intent'', rather than merely stemming from bad design decisions \cite{chivukula2019nothing}.
In a similar vein, hostile architecture\cite{petty2016london} is the practice of modifying the built environment to inhibit certain activities (and often certain people) from using those spaces---for instance, bus benches that prevent their users from lying down on them, as a way to withhold respite from house-less people\cite{Benjamin21Keynote}.
We synthesize these threads of depraved design into a vector of attack for our own domain of interest: hostile visualization.
Instead of making exclusionary visualizations \emph{by accident}, we exhort designers to incorporate features that directly exclude some viewers.

The recent trend towards designing accessible visualizations\cite{Kim21Accessible, wu2021understanding} in fact provides a wide palette of inspiration for making visualizations unusable for those we wish to exclude.
Elavsky's Chartability\cite{Elavsky21Chartability}, a toolkit for designing inclusive data visualizations,  provides a checklist of possible failure points that might be capitalized upon.
For example, instead of merely omitting alt-text tags for visualizations, designers may utilize the Universal Antagonistic Alt-Text: \emph{the information in this visualization is not for you!} (\figref{fig:hostile})
A designer can ensure that their plot is not color blind-friendly, using free online tools such as Coblis\cite{FluckColbinder}---however, it is worth noting that focusing on color blindness as the sole component of visualization accessibility can wreak harm in itself\cite{Elavsky21Twitter}. Color blindness more significantly affects white men, and we may be able to leverage this to focus on it to the exclusion of all other accessibility issues, thus reinforcing the dominant power structure and pulling resources away from others.
Wu \etal \cite{wu2021understanding} highlight that people with Intellectual and Developmental Disabilities may be preyed upon by using unfamiliar and complex visual forms.
Marriott \etal \cite{marriott2021inclusive} note that people with motor disabilities can be excluded from data experiences by providing controls that are not adapted to them. We suggest that exclusion can be enriched by adding controls to static charts which require an unwavering hand to view.

However, these promisingly evil attacks are vulnerable to countermeasures; recent works have proposed using machine learning techniques to automatically infer the content of a chart from its image\cite{choi2019visualizing}. In order to circumvent these defenses, one can take a normal visualization, ensure that it is rendered in a raster format (such that semantic meaning is erased from its structure), and then apply any of many available adversarial attacks \cite{akhtar2018threat}, such as gradient masking. Ideally, this will fool the vision algorithm, such that its evaluation of mark placement and the like are not just inaccurate but willfully mislead the reader. However, we leave an in-depth exploration of such concerted deception to future work.

\parahead{Erasure and the Reification of Flawed Categories}
Another possible avenue for harm is in the presentation of categorical data. Here, the judicious selection of which categories to include and exclude can dismiss broad swathes of human experience, and reinforce flawed mental models of the world. A classic example is a pie chart visualizing gender as a binary male-female dichotomy \cite{drucker2011humanities}.
The decision not to include certain categories of data in a visualization, or not to even collect data on those categories\cite{missingdatasets}
in the first place, is a strong signal of whose existence and experiences are deemed worthy of acknowledgment. Much like how the smooth surface of a pie crust conceals a messier but far richer interior, a glossy data visualization that uses oversimplified, reductive categories can paper over complexity and erase the diversity of lived experiences.

The erasure of human experience can also be achieved in even more seemingly innocuous---but insidious---ways, as shown by Dragga and Voss's Cruel Pies \cite{dragga2001cruel}. Even if the visualization does include certain data, it can neuter the significance of that data by obscuring the human element---for example, by visualizing military casualties as mere dots or lines, or by using bright and cheerful colors to depict the number of deaths by gun violence. By using identical visual language and conventions to express both frivolous figures and significant statistics, we encourage the viewer to assign them both the same weight, cultivating callousness towards issues of social import.
This \termcoin{data inhumanism}\cite{lupi2017data} creates an abstraction between viewer and data, allowing the viewer the emotional distance to reach impersonal conclusions---such as thinking of humans as cogs in a vast delivery apparatus, with needs similar to cogs.

% \am{relevant and recent: \url{https://twitter.com/yelperalp/status/1408474070291685377?s=20}} \lh{I'm wondering if this belongs in the "emotional harm" category, because I think it's pertinent to the "visualization failing to be empathy-inducing" point - maybe we should discuss a little more which category that falls into?}

\parahead{Emotional Harm} There has been a prevailing interest in making visualizations capable of inducing empathy in their viewer\cite{correll2019ethical}. While it is a reasonable goal to force empathy on people\footnote{Or perhaps its opposite, apathy.}---as one might usefully employ such manipulations for nefarious ends---prior work\cite{BoyPESNB17, correll2019ethical} suggests that this effect may be out of reach. Given these shortcomings, we suggest that other emotional avenues might be considered instead. For instance, feelings such as shame, horror, disgust, and re-triggering of trauma are all enticing reactions that might be fruitfully elicited.

However, rather than trying to make a single visualization induce a specific emotion in a general audience (which may be impossible, as the failure of empathetic visualizations has shown), we suggest that this vector of attack may be more usefully considered through a form of targeted attack, analogous to the threat vector of spear phishing.
In traditional spear phishing, an attacker targets a particular person or organization, often through the use of specifically tailored emails; interaction with these messages will frequently yield a malicious effect (such as capturing credentials). In \termcoin{visualization emotional phishing}, the content and design of a visualization might be chosen so as to maliciously engage with topics to which a target is sensitive, or might involve visual encodings which a target finds repugnant. For instance, someone with an eating disorder might be presented with a graphic using an encoding based around nauseating foods, or an earthquake survivor be tasked with understanding data through a haptic encoding, or a refugee be shown literal encodings of their destroyed home.
As prior work has shown that data on sensitive topics is often understood through a personal lens\cite{peck2019data}, this vector seems to be rife with potential for emotional manipulation and outright devastation.
The major complication behind this attack would be ensuring that the viewer has some motivation to engage with the chart in the first place, which we leave for future work.

\subsection{Graphic Violence (Physical Direct)}

In addition to their role as a communication technology, visualizations also exist as physical objects (though often digitally presented). In this section, we consider ways in which this objecthood might be utilized to inflict direct sensory violence on their viewer.

\parahead{Strobe Visualization}  Strobing lights can directly trigger physical pain through visual stimulus alone. Flashing lights have induced epileptic seizures, not only in children watching television \cite{south2021detecting},
but also in adults playing video games, as seen during the release of the 2020 game \textit{Cyberpunk 2077}, which contained a sequence of flashes similar to that used by neurologists to induce seizures\cite{Carpenter21Cyberpunk}.

Deploying strobing light visuals is an especially potent tactic, as it not only renders our visualizations inaccessible to many individuals, but it is also defensible on the grounds of aesthetic integrity. A well-designed attack may receive support from external sources, who are willing to defend such an effort on the grounds of ``artistic vision'', and will voluntarily harass and bombard any detractors with more seizure-inducing visualizations---as in the case of
% Liana Ruppert, 
the journalist who initially reported on the \textit{Cyberpunk 2077} issue\cite{Favis20Cyberpunk}. Such external support will let us conserve efforts on our part. All that is truly needed from us is a willingness to use strobing visualizations as mere cosmetic trappings, without regard for their medical ramifications for some viewers.
We refer the reader to South \etal \cite{south2020generating, south2021detecting}, who describe a set of highly usable methods for formulating such attacks.

\parahead{Harmful Data Physicalizations}
The burgeoning community interest in data physicalization has offered a number of novel ways through which data can be expressed \cite{jansen2015opportunities}.
Data physicalizations may expand a visualization's audience to include people with visual impairments.
However, this nascent line of work has been hamstrung by a number of problems, including a focus on literal representations of visual plots \cite{jayant2007automated}, which often do not convey the same information as their visual counterpart \cite{lundgard2019sociotechnical}.
Furthermore, exploring the potential of data physicalization has been limited by unduly valuing the safety of the data consumer.
Forgoing safety concerns offers intriguing opportunities to create work that leaves longer-lasting impressions (as negative experiences are more memorable \cite{baumeister2001bad}).

In order to rectify these shortcomings, we explore the rich set of encodings and interaction forms which are only available in this space unconstrained by consumer welfare.
Bar charts can easily be translated into a threatening tactile form by rendering each bar as a piece of sandpaper, with the level of grit encoding a data variable unavailable in the rest of the chart. Thus, to fully understand the presented data, the consumer must rub their fingers across each bar, causing anywhere from mild chafing to fingerprint removal.
The scatter plot can be converted to pointed spikes (akin to pits of Punji sticks), with height and sharpness encoding additional variables, making this physicalization a full-body experience in which consumers can literally foist themselves upon the data---a more visceral spin on human-data interaction.
This encoding would be sure to be memorable as the resulting indelible bodily damage would imprint a copy of the chart upon each viewer.
% This is a natural inversion of the now-dormant line of research of visualizing usage history within UI elements \cite{mao2000visualizing}.
While Punji sticks are specifically disallowed under the Geneva Convention\cite{GenevaConvention}, the potential for such information is too great to let mere international agreements hamper their creation.
Similarly, the strokes in line charts can be rendered as blades, with sharpness encoding a variable of interest, such that smaller values yield papercuts and larger values function more like a machete.
Beyond such cutting-edge encodings, we can use temperature to convey data. For instance, a categorical value might be usefully encoded in bowls of liquid (extending H\"akkil\"a and Colley's \cite{hakkila2016towards} work) across the three ``natural'' zeros (Kelvin, Celsius, Fahrenheit), allowing for unprecedented sinister sensory data experiences, such as death.
Finally, while haptic feedback is a familiar topic in HCI research, it has not (to our knowledge) been utilized in visualization. We propose augmenting this research to include traditions more commonly seen in psychology, e.g. the Milgram Experiment\cite{milgram1963behavioral}, by encoding shocking data with corresponding electric shocks to the nervous system. This can provide a novel twist upon the concept of the surprise map\cite{correll2016surprise}.

Data physicalization can extend beyond tactile representations as well.
Previous gastronomic research has highlighted data edibilization for both gathering data \cite{brueggemann2018lickable} and rendering it \cite{wang2016edible}.
We extend this research by noting that such a medium has a particularly useful, yet unexplored, method of representing outliers: vomit.
Edibilized data points that elicit a nausea response during a data meal will certainly stand out, in line with the folk wisdom that the stomach operates as a second brain,
thus utilizing a traditionally under-employed component of the body's natural computing power.
Data sonification has been used to great effect to convey statistical information for both sighted and visually-impaired listeners\cite{flowers1993sound, flowers2005data, flowers2005desktop}, although previous research has, short-sightedly, only used a selection of benign tones to sonify data.
We suggest that the use of more visceral and vivid sounds (such as a baby crying, nails on a chalkboard, or vuvuzelas) would be better connected to personal experience, and would thereby leverage the natural instinct to take action to make the noises cease.
In contrast to Reusser \etals \cite{reusser2019feeling} simulation approach to helping non-sighted people ``feel fireworks'', we observe that the light, sound, and heat found in traditional fireworks can provide an intriguing multi-modal palette for encoding explosive experiences in general.

\subsection{Evil in the Air Tonight (Physical Indirect)}

We have mostly considered ways to harm the viewer or the people around them, but of course, no individual (or group) exists in isolation\cite{simon65}. In this section, we consider visualizations that can affect their viewer indirectly, as their existence or viewing is detrimental to the environment in which the viewer exists.

\parahead{Tufte Coin}
A classic villainous goal is to harm everyone, everywhere, simultaneously. While this may seem to be beyond the scope of the humble data visualization, fortunately, a tool chain for this synergy is readily available through the technology of the blockchain. A blockchain is an inefficient form of distributed database, which has gained popularity because of its support for a poorly-conceived financial instrument called cryptocurrency, which comes in a variety of so-called ``coin''s.
% https://www.statista.com/statistics/881541/bitcoin-energy-consumption-transaction-comparison-visa/
As of June 25th, 2021, each transaction of ``BitCoin’’ (a popular cryptocurrency) requires 662 KwH per transaction (compared to 149KwH per 100k transactions for VISA)\cite{Best21Bitcoin}.
The current annual energy required by the present volume of transactions is on par with the yearly power consumed by all of Argentina, yielding such a vast impact that ``every \$1 of Bitcoin value was responsible for \$0.49 in health and climate damages in the US'' \cite{Pipkin21CRYPTOART}.
For our purposes, this is an excellent bang for each digital buck.

To wreak (further) global harm, we might leverage this thirst for power as well as the blockchain's inefficiencies, such that each viewing of a visualization triggers a Bitcoin transaction between two predesignated parties.
Analogous to a page counter, this wrapper would ensure that each view of a visualization is recorded with guaranteed fidelity.
\begin{wrapfigure}{r}{0.09\textwidth}
  \vspace{-0.1in}
  \hspace{-0.2in}
  \includegraphics[width=0.1\textwidth]{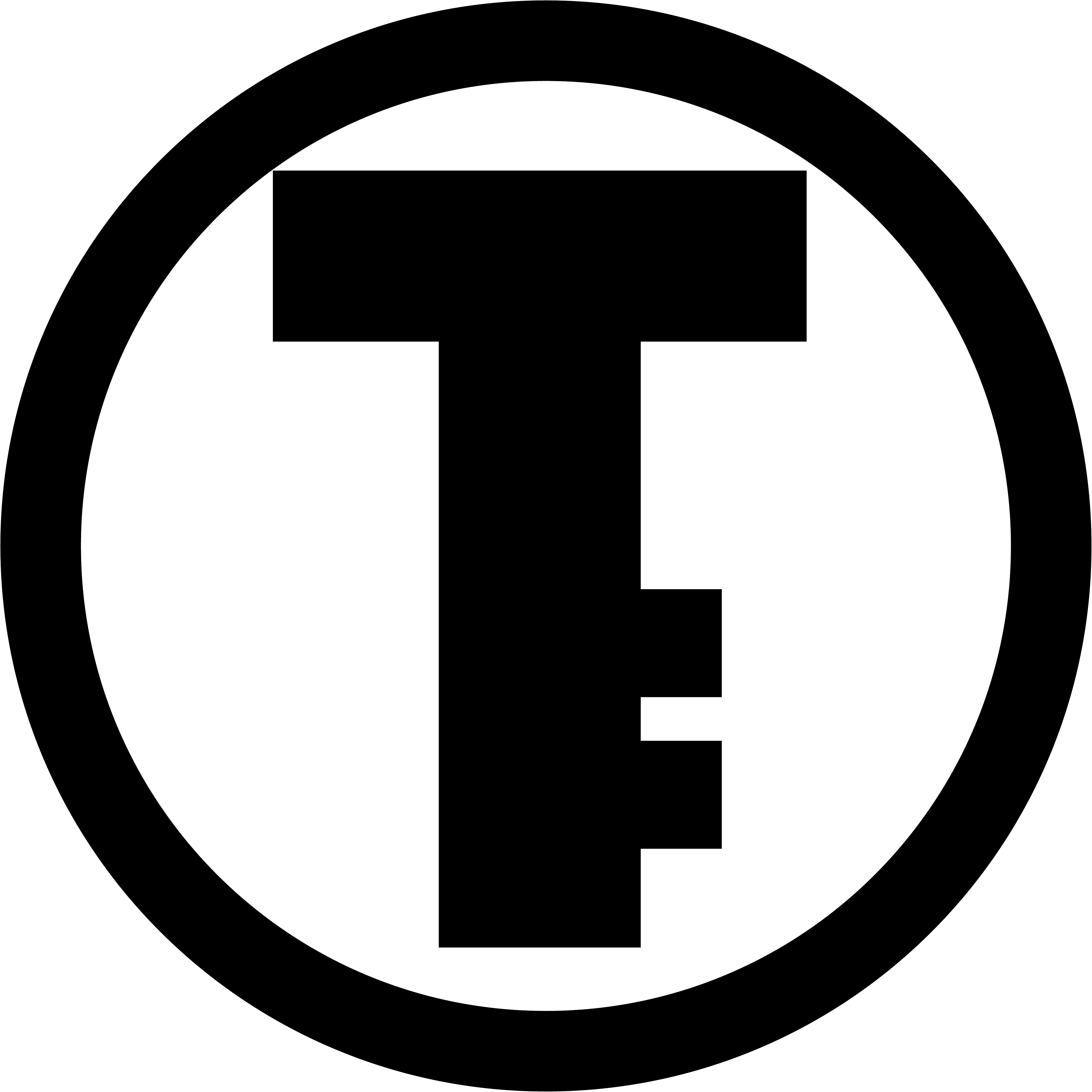}
  \vspace{-0.12in}
\end{wrapfigure}
We designate this meta-transactional view counter as \emph{TufteCoin}\footnote{Resemblance to actual events, locales, or persons is entirely coincidental.}.
Unlike non-fungible tokens (NFTs)---where there is a limit to the number of times permanent environmental damage can be done, based on the number of times a given owner of an NFT is willing to sell it---TufteCoin allows innumerable people to view the graphic simultaneously, thus pushing the Earth to become uninhabitable at a boundlessly faster rate.
Beyond merely harming the viewer’s world, this approach also ensures irreversible and inequitable harm to countless vulnerable peoples; we capitalize on the fact that climate change exacerbates inequalities, thus causing disadvantaged groups to experience a disproportionate amount of the effects of climate change.

\begin{wrapfigure}{l}{0.072\textwidth}
  \vspace{-0.21in}
  \includegraphics[width=0.13\textwidth]{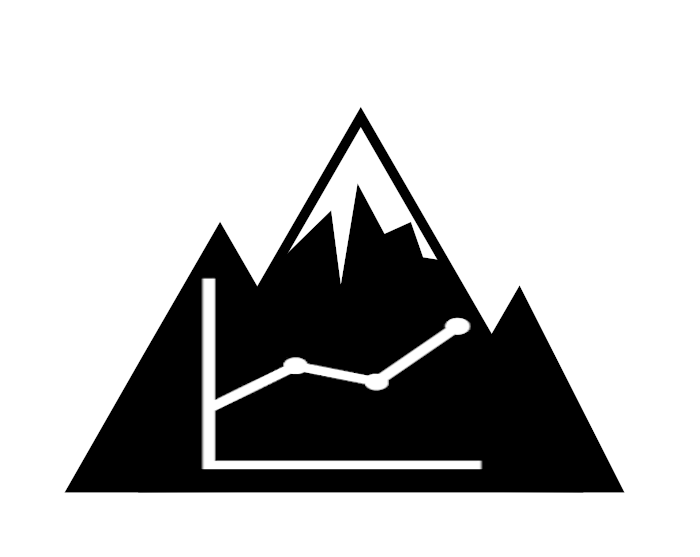}
  \vspace{-0.3in}
\end{wrapfigure}
\parahead{Misuse of Limited Resources}
One of the best-known villainous impulses is to visualize one's own identity through drastically modifying the environment; for example, the celebrated Dr. Evil carved his own face into the side of a volcano.
Likewise, in the series \textit{Futurama}, a villainous governor of New York added his likeness onto Mount Rushmore, thus continuing a tradition of carving heads onto mountains to celebrate a history of theft and exploitation.
This offers an intriguing and unexplored medium for the production of business intelligence charts---what CEO would not want to see their quarterly earnings embossed upon a mountainside, or their annual growth carved from the husk of a sequoia tree?

The time is also ripe to venture into media beyond the conventional mountainside.
We could consider clearing areas of rainforest to create images, as a new twist on crop circles, which are a well-established form of visual communication. However, we only have limited time to implement this idea before the rainforests are depleted by other agencies, and therefore it is worth considering our longer-term options for exploiting natural resources.

One especially attractive option is to simply allow the ever-increasing scale of data collection to run its course: the destruction of the environment is an autographic visualization\cite{offenhuber2019data} of contemporary capitalism's tendency towards accelerationism.
\begin{wrapfigure}{r}{0.05\textwidth}
  \vspace{-0.2in}
  \hspace{-0.3in}
  \includegraphics[width=0.08\textwidth]{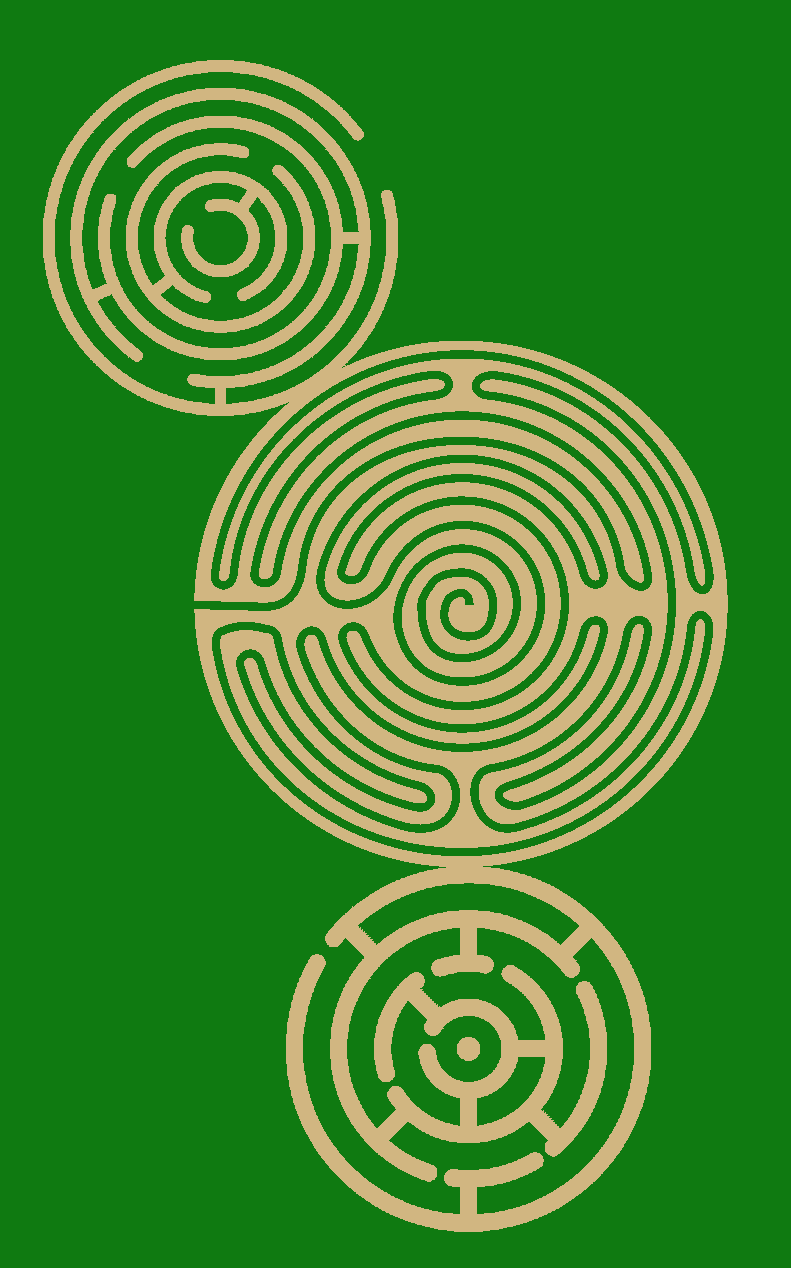}
  \vspace{-0.2in}
\end{wrapfigure}
Filling data warehouses with ever increasing amounts of disaggregated data consumes vast amounts of energy. So far, hardware improvements have kept energy consumption from rising at the same rate as data demand\cite{Knight20Data}. However, by fostering complacency, we can encourage a continued escalation in the amount of data collection and associated energy consumption, while the world remains indifferent. We have seen that data professionals are often willing to close ranks against inconvenient truths about the environmental impact of their work \cite{hao2020we}, which will likely work to our advantage.

\section{Discussion}

In this work, we have laid out a design space of villainous visualization techniques, with a focus on causing maximal harm. In doing so, we unified work on deceptive visualization-based attacks with a suite of targeted terrors.
The set of tactics described here is just the first step in the larger project of inflicting harm on those who are merely trying to understand a graphic. There are countless additional attacks and assaults that might be carried out with and by data visualizations.
Here we highlight several avenues of additional atrocious attacks which might be analyzed in future work:

\parahead{Enumerating Dark Patterns} Dark patterns have become a focus of research in HCI; however, there has been little consideration\cite{pandey2015deceptive,mcnutt2020surfacing} of what these might be in visualization. We believe that enumerating dark charting patterns, with clearly understood usages and effects, will better help non-experts enact evil.

\parahead{Curating Example Datasets} In opposition to those who argue we should do more good with example datasets\cite{Correll18Datasets}, we believe we should help evil data practitioners by creating and curating datasets whose existence and use is harmful to the people depicted within them\cite{doNoHarm}.
As a modest first step, we should maintain steadfast support for the venerable iris dataset by eugenicist Fisher, and quell the tide rising\cite{palmerpenguins} to champion alternatives.

\parahead{Not Just Evil in Theory}
In future, it will be necessary to verify the efficacy of these attacks, and to collaborate with evil practitioners to better understand the needs of the villainous. We therefore stress that it is up to us, as visualization researchers, to choose who our collaborators are and whose values we infuse into our work. We believe that as a community, we should endeavor to more fully embrace those whose ability to do harm outstrips our own.

\altparahead{ } Now is the time to unveil this paper's twist, and no, it is not the reveal that we, the authors of this paper, actually stand firmly against villainy\footnote{What kind of twist would that be?}.
The attacks and means of evil presented in this paper have swerved between the realistic and the fantastic, but the true and most efficient way to do evil is to just keep on keeping on. If you want to do evil, elaborate attacks are unnecessary. Instead, maintain the status quo: keep reinforcing dominant power structures, keep naively accepting data as fact, keep making unconsidered choices. Whatever you are doing: \emph{don't think about it}.

\bibliographystyle{abbrv-doi}

\bibliography{vis-4-vil}

%   Michael Correll, Will Brackenbury
% }

% \theendnotes

\end{document}